\documentclass[useAMS,usenatbib]{mn2e}

\title[$\alpha$ Centauri A and B]{Models of $\alpha$ Centauri A and B with and without seismic constraints: 
time dependence of the mixing-length parameter}
\author[M. Y{\i}ld{\i}z]{M. Y{\i}ld{\i}z$^{}$\thanks{E-mail:
mutlu.yildiz@ege.edu.tr},\\
Ege University, Department of Astronomy and Space Sciences, Bornova, 35100 \.Izmir, Turkey}

\input{psfig}

\begin{document}

\date{Accepted 2005 December 15. Received 2005 December 14; in original form 2005 October 11}

\pagerange{\pageref{firstpage}--\pageref{lastpage}} \pubyear{2006}

\maketitle

\label{firstpage}

\begin{abstract}
{ %
The $\alpha$   Cen binary system is a well-known stellar system with very
accurate observational constraints to structure of its component stars.
In addition to the classical non-seismic constraints, there are also seismic
constraints for the interior models of $\alpha$   Cen A and B. These two types of 
constraint give very different values for the age of the system. While we obtain
8.9 Gyr for the age of the system from the  non-seismic constraints, the seismic
constraints imply that the age is about 5.6-5.9 Gyr. There may be observational or theoretical reasons for this discrepancy,
which can be found by careful consideration
of similar stars. The $\alpha$   Cen binary system, with
its solar type components, is also suitable for testing the stellar mass
dependence of the mixing-length parameter for convection derived from the
binaries of Hyades. The values of the mixing-length parameter for $\alpha$  Cen A and B
are 2.10 and 1.90 for the non-seismic constraints. If we prioritize to the seismic
constraints, we obtain 1.64 and 1.91
for $\alpha$ Cen A and B, respectively. By taking into account of these two contrasting cases 
for stellar mass dependence
of the mixing-length parameter, we derive two expressions for its time dependence,
which are also compatible with the mass dependence of the mixing-length parameter
derived from the Hyades stars.
For assessment, these expressions should be tested in other stellar systems and clusters. 
}
%
\end{abstract}


\section{Introduction}
The observable quantities of a star, such as luminosity, radius and effective temperature,
mostly depend on its mass. Chemical composition, rotation and age, however, have a second role in these 
quantities.  The time dependence of the observable quantities are very weak. 
Therefore, no more or less precise age for stars available, particularly for late-type stars, and this is 
an essential quantity for our understanding of the evolution of both the far and near Universe,
However, it is possible to derive the ages of stars from their seismic properties (Christensen-Dalsgaard, 1988).
In this respect,  the solar-like components of $\alpha$ Centauri ($\alpha$ Cen)
are excellent targets, as their seismic properties are determined by ground-based spectrographic 
observations. The aim of the present paper is to test the theory of stellar evolution by constructing 
models for the internal structure of these stars, using all the available constraints. 

The internal structures of $\alpha$ Cen A and B 
have been considered theoretically in many papers. Recent studies by
Miglio \& Montalban (2005), Eggenberger et al. (2004),  Thoul et al. (2003), Morel et al. (2000) and Guenther \& Demarque
(2000) have analysed the structure and evolution of these stars in detail.
The detection of their seismic properties have increased
the scientific interest in these stars.   
The large and small frequency separations of $\alpha$ Cen A and B have been 
derived observationally from their p-mode oscillations by Bouchy \& Carrier (2002),  
and Carrier \& Bourban (2003), respectively, using spectrographic methods. In addition to the constraints from the fundamental properties of 
the stars, the interior models of these stars should also satisfy these seismic constraints.
Recently,  Kjeldsen et al. (2005) also observed
$\alpha$ Cen B, and measured the frequencies, which turned out to be quite different from the previous results of Carrier \& Bourban (2003). 
Therefore, the system deserves to be reconsidered by taking  this new measurement into account.

The masses (Pourbaix et al. 2002) and radii (Kervella et al. 2003) of $\alpha$ Cen are well known from the 
cited observations.
Although the $\alpha$ Cen system has already been widely investigated, no agreement has yet been 
reached about the precise value of the effective temperatures ($T_{\rm eff}$) and element abundances in the
atmospheres of these stars. These quantities are determined by spectroscopic methods. However, 
using the photometric observation of Bessel (1990) as an alternative, which has unfortunately
been neglected in studies on $\alpha$ Cen, we have alternatively derived the effective temperatures and the 
heavy element abundances of both components.

{%
The effective temperature of a model is a function of the 
}
mixing-length parameter for convection.
Y{\i}ld{\i}z et al. (2006) have confirmed a very definite
relationship between the mixing-length parameter and the stellar mass by fitting the models of late type components 
of some binaries in Hyades to the observations:  $\alpha = 9.19 (M/M_\odot-0.74)^{0.053}-6.65$. 
However, controversial results are obtained from 
studies on binaries by different investigators: while Lastennet et al. (2003), 
Ludwig \& Salaris (1999), and Lebreton et al. (2001) have stated that  $\alpha$ is an 
increasing function of the stellar mass in their studies on some other binaries, 
Miglio \& Montalban (2005), Eggenberger et al. (2004), Morel et al. (2000) have confirmed that $\alpha_{\rm B}$ is greater than
$\alpha_{\rm A}$ for the $\alpha$ Cen binary system. In these studies and others, at least for the favoured models, however,
we note that the model luminosity is less than the
observed value for $\alpha$ Cen A,  and that the opposite is true for $\alpha$ Cen B.
This point needs to be explained.
However, the results of the hydrodynamic simulations of convection
are in agreement with the decrease  $\alpha $. Thus, there may be no single relationship 
between the mixing-length parameter 
and the stellar mass and we should therefore search 
{%
for a  more general expression for $\alpha$ 
}
than that given above.

%



According to the above-mentioned studies on the internal structure of $\alpha$ Cen A and B, the age of the system 
is between 4.85-7.6 Gyr, at least for the favoured or typical models. The lower limit is 
given by Thevenin et al. (2002) and
the upper limit is from Guenther \& Demarque (2000). Miglio \& Montalban (2005), however, find the age of the 
system {to be} 8.9 Gyr from the models for the non-seismic constraints (see Section 4.1). In other words,
the seismic and  the non-seismic constraints lead to very different ages for $\alpha$ Cen.


The remainder of this paper is organized as follows. In Section 2, we summarize the observed chemical composition, the seismic and the fundamental properties
of $\alpha$ Cen A and B.  
The description of models are given in Section 3, and the results of the model computations are presented and discussed in
Section 4.
Finally, we give 
concluding remarks in Section 5.

\section[]{Observed properties of $\alpha$ Cen}

\subsection[]{Chemical abundance of $\alpha$ Cen A and B}
Many studies have been done on the abundance determination of chemical species of $\alpha$ Cen from 
the spectra of its components. Most of these studies 
deal with the abundance of Fe and/or similar heavy elements, which are not abundant in stars. The most complete studies--in the sense that
they also deal with the most abundant heavy elements in $normal$ stars such as oxygen, carbon and nitrogen--have been cwcarried out by 
Neuforge-Verheecke \& Magain (1997) and Feltzing \& Gilmore (2001, hereafter FG2001). The abundances of 17 elements have been obtained by
Neuforge-Verheecke \& Magain (1997). 
They find from the spectra of $\alpha$ Cen A that overabundance relative to solar is 0.25 dex for Fe and  
0.21 for the most abundant heavy element oxygen, and a general overabundance of 0.24 dex. They also confirm that there 
is no significant difference between the chemical compositions of $\alpha$ Cen A and B, but $\alpha$ Cen B { is} likely more metal rich 
than $\alpha$ Cen A.  Using their equivalent widths, 
FG2001 have obtained very similar results for the abundance of 13 heavy elements, including oxygen.

By considering Mg, Ca, Si, Ti, Cr and Fe elements, Doyle et al. (2004) have found that the mean 
overabundance of $\alpha$ Cen A is 0.12$\pm$0.06 dex. Contrary to this result, Ecuvillon et al. (2006) computed 
the average value of the oxygen abundance of $\alpha$ Cen A and B as [O/H]=-0.12 and -0.06, 
respectively, from triplet lines by applying non-local thermodynamical equilibrium effects. Such contradictions in the abundance determination $\alpha$ Cen A and B, and  
the recent revisions on the solar compositions (Asplund et al. 2005) make us cautious about  seeing the abundance 
determination of stars as an issue that has been conclusively studied.

{The mixture in FG2001 is taken as the chemical composition of the OPAL opacity tables 
(Iglesias \& Rogers 1996) used in the construction of models for $\alpha$ Cen A and B.}
For the chemical species that are not observed by FG2001, such as Ne, 
we adopt the value of  0.25 dex for their abundance relative to the Sun. For comparison, we also construct models with the recent solar mixture given by Asplund et al. (2005, hereafter AGS2005).

\subsection[]{Fundamental properties of $\alpha$ Cen A and B}

\begin{table*}
\caption{
 The fundamental properties of $\alpha$ Cen A and B. Using visual magnitudes (Hoffleit \& Jaschek 1982; Bessel 1990) and
     the observed masses and radii of $\alpha$ Cen A and B,  we find
     bolometric corrections from the tables of LCB1998 and Bes1998, and then their luminosity for selected values of Z. The results given in the last two rows are obtained
     by also fitting the  colours of $\alpha$ Cen A and B. In this way, we also determine the current heavy element abundance of each star.  
}
\label{ta1}
$
\begin{array}{p{0.15\linewidth}cccccccccl}
\hline 
\hline
            \noalign{\smallskip}
Star      & L_{\rm A}/L_\odot& T_{\rm eff} (K) & M_{\rm Bol}& BC &   U-B & B-V & M_{\rm V} & \left[Z/Z_\odot \right] &Table(BC)\\
            \noalign{\smallskip}
            \hline
            \noalign{\smallskip}

$\alpha$ Cen A  & 1.552& 5831& 4.2728& -0.0852&  0.2042&  0.6583&  4.3579& 0.25dex& LCB1998\\
$\alpha$ Cen A  & 1.555& 5834& 4.2707& -0.0876&  0.1837&  0.6509&  4.3583& 0.20dex& LCB1998\\
$\alpha$ Cen A  & 1.585& 5862& 4.2499& -0.1081&  0.0721&  0.6077&  4.3580& 0.00dex& LCB1998\\
$\alpha$ Cen A  & 1.520& 5801& 4.2954& -0.0623&  0.1578&  0.6642&  4.3577& 0.00dex& Bes1998\\
            \hline
$\alpha$ Cen B  & 0.511& 5260& 5.4796& -0.2174&  0.5922&  0.8559&  5.6969& 0.25dex& LCB1998\\
$\alpha$ Cen B  & 0.511& 5261& 5.4789& -0.2176&  0.5690&  0.8486&  5.6966& 0.20dex& LCB1998\\
$\alpha$ Cen B  & 0.515& 5271& 5.4705& -0.2261&  0.4258&  0.8037&  5.6966& 0.00dex& LCB1998\\
$\alpha$ Cen B  & 0.497& 5225& 5.5082& -0.1885&  0.4746&  0.8552&  5.6968& 0.00dex& Bes1998\\
            \hline
$\alpha$ Cen A  & 1.544 &5824& 4.2784& -0.1019&  0.1276&  0.6333&  4.3803&0.074dex& LCB1998\\
$\alpha$ Cen B  & 0.507 &5250& 5.4875& -0.2227&  0.5349&  0.8398&  5.7102&0.125dex& LCB1998\\
            \noalign{\smallskip}
            \hline
\end{array}
$
\end{table*}
For our better understanding of stellar structure and evolution, a very accurate knowledge of 
stellar masses is required. The determination of masses of $\alpha$ Cen A and B from observations 
has been the subject of many papers. The most recent determination is that of Pourbaix et al. (2002). According to their findings,
the masses of $\alpha$ Cen A and B  are as 
$M_{\rm A}=1.105 \pm 0.007$ M$_\odot$ and $M_{\rm B}=0.934\pm 0.007$ M$_\odot$, respectively. The radii of 
the components have been measured by Kervella et al. (2003) as 
$R_{\rm A}=1.224 \pm 0.003$ R$_\odot$ and $R_{\rm B}=0.863 \pm 0.005$ R$_\odot$.
The uncertainties in radii are computed by taking into account only the uncertainty in the parallax.
Many investigators have determined the effective temperatures from the spectra of $\alpha$ Cen A and B is done by many investigators
(see, for example, Morel et al. 2000 and references therein). 
According to the results of studies based on spectra,  
while the effective temperature of $\alpha$ Cen A is between 5830 K and 5720 K, { the} range of effective 
temperatures of $\alpha$ Cen B is 5250-5325 K. 

In the literature, apparent magnitude (V) and colour (B-V) of $\alpha$ Cen A and B are, in general, taken from  
The Bright Star Catalogue (Hoffleit \& Jaschek 1982): $V_{\rm A}=-0.01$, $V_{\rm B}= 1.33$, $(B-V)_{\rm A}= 0.71$  
and $(B-V)_{\rm B}= 0.88$. Adopting the parallax given  in S\"oderhjelm (1999), $\pi=747.1\pm 1.2$ 
mas, we find
the absolute magnitudes of $\alpha$ Cen A and B as 4.358 and 5.697, respectively. Using the tables of 
Lejeune et al. (1998, hereafter LCB1998) 
for colours and bolometric correction, prepared for different heavy element abundances, we obtain the required luminosities 
of $\alpha$ Cen A and B for selected values of Z. In Table 1 the first three rows for $\alpha$ Cen A and B are for the heavy element abundance of 
0.25 dex and 0.20 dex and for the solar metallicity, respectively. For each metal abundance, while we obtain reasonable values for the 
absolute magnitudes from the tables in comparison with the observed values given above, for the colours, there is no 
agreement between theoretical and observational results.  In the fourth and the eighth rows, we derive the fundamental properties 
{ 
of $\alpha$ Cen A and B, respectively,
}
using the tables of  Bessel, Castelli \& Plez (1998, hereafter Bes1998) for colour and bolometric correction, prepared with the solar composition.   
While the luminosities of $\alpha$ Cen A and B found from Bes1998 are smaller than the corresponding luminosities 
computed from the tables of LCB1998, the colours  computed from the two tables are significantly less than the observed colours for all the selected 
metallicities.  

In Bessel (1990), however, the apparent magnitudes and the colours of (B-V) $\alpha$ Cen A and B are given as
 $V_{\rm A}=0.01$, $V_{\rm B}= 1.34$, $(B-V)_{\rm A}= 0.633$
and $(B-V)_{\rm B}= 0.84$. 
{
These apparent magnitudes give absolute magnitudes of
}
 $M_{\rm V, A}=4.377$ and $M_{\rm V, B}=5.707$.
Using the above mass, radius, $M_{\rm V}$ and $B-V$ of each component we derive the luminosity, effective temperature
and also the current value of the surface heavy element abundance ($Z_{\rm S}$) from the tables of LCB1998.
Their tables are prepared 
{  
with different values of $Z$,
}
which enable us to determine 
$Z_{\rm S}$. By changing $Z$ and the luminosity of the component stars, we obtain 
the observed absolute magnitude and (B-V) of each star. 
The fundamental properties of $\alpha$ Cen A and B 
we have derived are listed in the last two rows of Table 1, respectively. The current values of the 
surface heavy element abundances of
$\alpha$ Cen A and B relative to the Sun are found as $0.074$ $ dex$ and $0.125$ $ dex$, respectively. 
The value of Z depends on what value is adopted for $Z_{\odot}$ in the tables of LCB1998. 
For $Z_{\odot}=0.02$, $Z_{\rm S,A}=0.0237$, $Z_{\rm S,B}=0.0267$. 
While the luminosities 
we find are slightly greater than those given by Eggenberger et al. (2004), the effective temperatures of
$\alpha$ Cen A and B are found as 5824 K and 5250 K, respectively, and { are} in very good agreement with the 
values found from spectra  by Neuforge-Verheecke \& Magain (1997). 

Indeed, 
the measurements given in Bessel (1990) and Hoffleit \& Jaschek (1982) are very close to each other; 
the remarkable differences { that} exist between the colours of each star and the { measurements of} Bessel (1990) are compatible with the 
theoretical results. Therefore, we adopt the photometric constraints in Bessel (1990) in the calibration { process }
of models for $\alpha$ Cen A and B.

\subsection[]{Seismic  properties of $\alpha$ Cen A and B and an age estimation for the system}
   \begin{figure}
\centerline{\psfig{figure=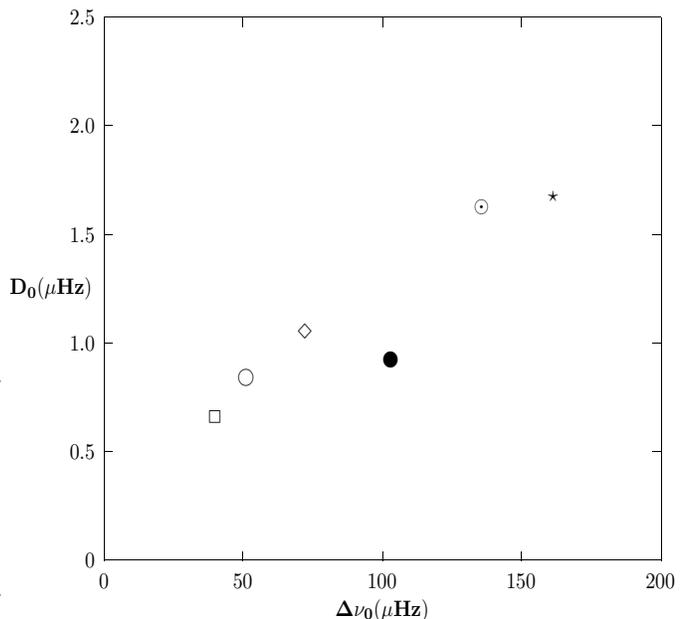,width=210bp,height=250bp}}
      \caption{Observational seismic HR diagram for the Sun ($\odot$), $\alpha$ Cen A (filled circle) and B (star), 
$\eta$ Boo (box), $\beta$ Vir (diamond) and Procyon A (circles). The seismic data are taken from Chaplin et al. (1999), Bouchy \& Carrier (2002), 
Kjeldsen et al. (2005), Carrier et al. (2005a,b) and Martic et al. (2004). 
$D_0$ is computed
from the small separations $\delta \nu_{02}$ between the observed frequencies. For the Sun and $\alpha$ Cen B, $\delta \nu_{13}$ 
is also available and is thus also used in the computation of $D_0$ of these stars.
}
              {\label{f1.1}}
   \end{figure}
The most important constraints on the stellar evolution theory from the observations of solar-like oscillations 
are small ($\delta \nu_{nl}$)  and large ($\Delta \nu_{nl}$) separations between the observed frequencies. 
The small separation between the frequencies (Christensen-Dalsgaard 1988) is defined as 
\begin{equation}
\delta \nu_{nl}=\nu_{n,l}-\nu_{n-1,l+2}.
\end{equation}
In place of this, $D_0$ is frequently used as an average effect of oscillations with 
different values of degree $l$. We compute $D_0$ 
from $\delta \nu_{n0}$: $D_0=\delta \nu_{n0}/6$ (Kjeldsen et al. 2005). If frequencies for $l=3$ modes are observed,
$D_0$ is computed from $\delta \nu_{n0}$ and $\delta \nu_{n1}$:$D_0=
{ 
(\delta \nu_{n0}/6+\delta \nu_{n1}/10)/2}$.  

The large separation between the frequencies is given below:
\begin{equation}
\Delta \nu_{nl}=\nu_{n,l}-\nu_{n-1,l}.
\end{equation}
While the small separation is a sensitive function of physical conditions in central regions where the nuclear evolution proceeds, the large 
separation is a measure of the mean density. 

The large and small frequency separations of $\alpha$ Cen A are determined by 
Bouchy \& Carrier (2002) as $105.5$ $\pm$ 0.1 $\mu$Hz and 5.6 $\pm$ 0.7 $\mu$Hz ($D_0=0.93\pm0.12 $ $\mu$Hz), respectively. 
Carrier \& Bourban (2003) find that $\delta \nu_{n0}=8.7\pm0.8$ $\mu$Hz and $\Delta \nu_{n0}=161.5\pm0.1$ $\mu$Hz for  $\alpha$ Cen B.
Recently, Kjeldsen et al. (2005) also obtained the seismic properties of $\alpha$ Cen B using
the two-site observations.
According to their findings, while the large separation between the frequencies is very similar to that of Carrier \& Bourban (2003),
 the small separation between the frequencies is $\delta \nu_{n0}=10.14\pm0.62$ $\mu$Hz and 
significantly larger than the previously found separation ($8.7\pm0.8 $ $\mu$Hz) by Carrier \& Bourban (2003).
Kjeldsen et al. (2005) also observed the oscillations with $l=3$ and the small separation between the frequencies of these 
oscillations and oscillations with $l=1$ is $\delta \nu_{n1}=16.73\pm0.65$ $\mu$Hz. From $\delta \nu_{n0}$ and $\delta \nu_{n1}$, $D_0=1.68\pm0.08 $ $\mu$Hz 
{
is found. 
} 

Fig. 1 shows and observational seismic HR diagram for several stars for which $D_0$ and $\Delta \nu_{nl}$ 
are available. While the filled circle and star represent $\alpha$ Cen A and B, respectively, $\odot$ shows the position of the Sun. $D_0$ and $\Delta \nu_{nl}$ of the Sun are computed from the data of Bison group (Chaplin et al. 1999). 
The seismic data of $\eta$ Boo (box), $\beta$ Vir (diamond) and Procyon A (circle) are taken from
Carrier, Eggenberger \& Bouchy (2005), Carrier et al. (2005), and Martic et al. (2004), respectively.
The $D_0$ values of $\alpha$ Cen B and the Sun are very close to  each other. This means that the central 
hydrogen abundance of $\alpha$ Cen B ($X_{\rm c}=0.34$) is nearly equal to the solar value 
(see figure 1 in Christensen-Dalsgaard 1988). Thus, the age of $\alpha$ Cen B is about half of 
its nuclear time scale ($t_{\rm nuc}$). From this confirmation, we can estimate the age of 
$\alpha$ Cen using a simple approach.  Because the  nuclear time-scale is inversely proportional to $M^{2.5}$ (B\"ohm-Vitense 1992),  
\begin{equation}
t_{\alpha \rm Cen}=\left(\frac{M_\odot}{M_{\rm B}}\right)^{2.5} t_{\odot}=5.5~{\rm Gyr}.
\end{equation}


The basic 
{
observational
}
properties of $\alpha$ Cen A and B discussed in this section are given in the last two rows of Table 2.  

\section[]{Description of The Models}
The characteristics of our code have  already
been described by Y{\i}ld{\i}z (2000, 2003, see also references therein),
and therefore we shall not go into details. 
Our equation of state uses the approach of Mihalas et al.~(1990) in the
computation of the partition functions. The radiative
opacity is derived from Iglesias \& Rogers (1996), completed
by the low temperature tables of Alexander \& Ferguson (1994).
For the nuclear
reaction rates, we use the analytic expressions given
by Caughlan \& Fowler (1988), and we employ
the standard mixing-length theory for convection (B\"{o}hm-Vitense 1958).
For the diffusion of chemical species, the routines of Thoul, Bahcall \& Loeb
 (1994) are used.

\section{Results and Discussions}
\subsection{Models satisfying the constraints on the fundamental properties of $\alpha$ Cen A and B}

\begin{table*}
\label{ta2}
\caption{
 Model properties of $\alpha$ Cen A and B. $\delta \nu_{02}$, $\delta \nu_{13}$, $ D_0 $ and $ \Delta \nu_0$ are  in units $\mu$Hz.
{
The uncertainty of $\delta \nu_{02}$ and $\delta \nu_{13}$ is about 0.7 $\mu$Hz. The error in $ \Delta \nu_0$ is 0.1 $\mu$Hz. The uncertainty of 
$ D_0 $ is 0.12 $\mu$Hz for $\alpha$ Cen A and 0.08 $\mu$Hz for $\alpha$ Cen B. 
}
}
$
\begin{array}{llllcclllllllllll}
\hline 
\hline
            \noalign{\smallskip}
Star&L_{\rm }/L_\odot& R_{\rm }/R_\odot & T_{\rm eff}& X_{\rm 0}& Z_{\rm 0} & Z_{\rm S} & (Z/X)_{\rm S} & \alpha & t(10^9 y) & \delta \nu_{02}&\delta \nu_{13}& D_0 & \Delta \nu_0& cc&MODEL\\
		    \noalign{\smallskip}
		    \hline
		    \noalign{\smallskip}

A  & 1.545& 1.224& 5822& 0.703& 0.0328&  0.0237& 0.0316& 2.10 & 8.88& 2.7& ......& 0.45&106.5& FG2001& NOS\\ 
B  & 0.507& 0.864& 5246& 0.703& 0.0328&  0.0267& 0.0364& 1.90 & 8.88& 6.6& 15.8  & 1.30&164.6& FG2001& NOS\\
    \hline
A  & 1.545& 1.224& 5822& 0.669& 0.0322&  0.0237& 0.0316& 1.64 & 5.70& 5.57& ......& 0.93&107.4& FG2001& SIS\\
B  & 0.584& 0.863& 5436& 0.669& 0.0322&  0.0271& 0.0388& 1.91 & 5.70& 9.15& 18.6  & 1.68&164.6& FG2001& SIS\\
    \hline
B  & 0.506& 0.862& 5248& 0.669& 0.0322&  0.0271& 0.0389& 1.58 & 5.70& 9.50& 19.0  & 1.74&163.5& FG2001&SIS914\\
 \hline
A  & 1.543& 1.223& 5822& 0.676& 0.03&  0.0198& 0.0316& 1.65 & 5.57& 5.64& ......& 0.94&106.8& FG2001& SIS3\\
B  & 0.584& 0.863& 5441& 0.676& 0.03&  0.0271& 0.0388& 1.93 & 5.57& 9.29& 18.4  & 1.70&164.6& FG2001& SIS3\\
    \hline
A  & 1.549& 1.223& 5828& 0.713& 0.023&  0.0161& 0.0211& 1.71 & 5.90& 5.49& ......& 0.91&106.7& FG2001& SIS23\\
B  & 0.569& 0.863& 5400& 0.713& 0.023&  0.0192& 0.0259& 1.98 & 5.90& 9.31& 18.5  & 1.70&164.4& FG2001& SIS23\\
    \hline
A  & 1.544& 1.223& 5828& 0.697& 0.0233&  0.0159& 0.0213& 1.67 & 5.70 & 5.61& ......& 0.93&106.8& AGS2005& SISAGS\\  
B  & 0.572& 0.863& 5409& 0.697& 0.0233&  0.0192& 0.0265& 1.96 & 5.70& 9.38& 18.6  & 1.71&165.0& AGS2005& SISAGS\\
		    \hline
		    \hline

A  & 1.544& 1.224& 5824& .... & .....&  0.0237^a & 0.0364^b & .... & ....&  5.6 & ...... & 0.93&105.5& .... & { obs}\\
B  & 0.507& 0.863& 5250& .... & .....&  0.0267^a & 0.0399^c & .... & ....& 10.14 & 16.73  & 1.68&161.5& .... & { obs}\\
		    \hline
		    \noalign{\smallskip}
		    \hline
	\end{array}
	$
{ 
~~~\\
$^a$Derived from the photometric data of Bessel (1990) using  the tables for colours  and bolometric corrections of LCB1998.\\
$^b$) Derived from abundances found by FG2001. For the elements not observed by them, an average overabundance of 0.25 dex relative to solar is assumed.\\
$^c$) Derived from $^b$, assuming an overabundance of 0.04 dex relative to $\alpha$ Cen B (observed for Al and Si, for example). 
}
	\end{table*}

	   \begin{figure}
	\centerline{\psfig{figure=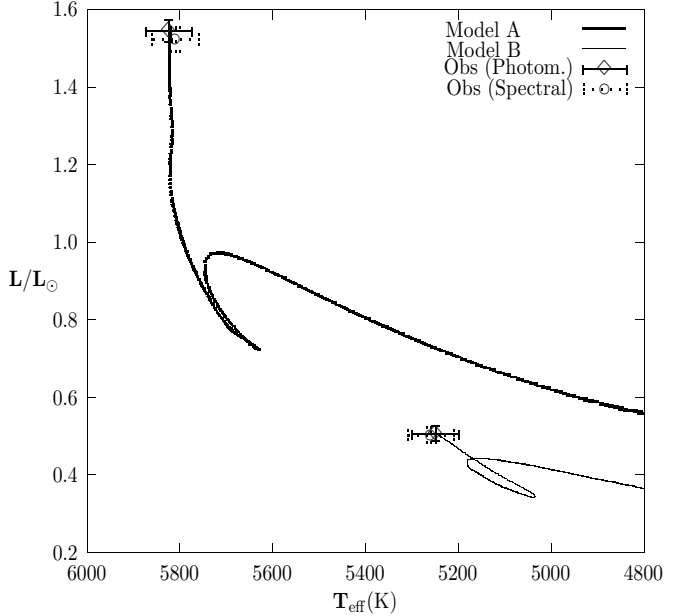,width=210bp,height=250bp}}
	      \caption{Evolutionary tracks of models (NOS model) for  $\alpha$ Cen A (thick line) and B (thin line) in the HR diagram.
		      While the diamond represents the data we derived from photometric data in Bessel (1990), the circle shows
		      the data given by Eggenberger et al. (2004)}
		      {\label{f1.10}}
	   \end{figure}

In order to find the five unknowns (i.e. X, Z, t, $\alpha_{\rm A}$ and $\alpha_{\rm B}$), we write down four equations for 
$L_{\rm A}$, $L_{\rm B}$, $R_{\rm A}$ and $R_{\rm B}$. The fifth equation for a solution is for the surface value of the (Z/X) ratio for $\alpha$ Cen A.
For $(Z/X)_{\rm A,s}=0.0237$, the values of five unknowns and the models of $\alpha$ Cen A and B using 
{ 
these values
} 
are given in the first two rows
of Table 2. The heavy element mixture of the models is taken from FG2001.  
According to the solution for non-seismic constraints (NOS model), { the} age of the system is 8.88 Gyr and $\alpha$ Cen A is 
very close to { the} end of its main-sequence (MS) lifetime. Because the seismic constraints are not used in this solution, the seismic properties of 
models are not in 
agreement with the observations: small separations between the frequencies of the models are significantly smaller
than the observed separations for both stars.  
In the last two rows of Table 2, the observational results are listed for comparison.
In Fig. 2 the models are plotted in a theoretical HR diagram. The observed positions of $\alpha$ Cen A and B derived from the 
photometric measurements (diamond) are also seen. For comparison, the data given by Eggenberger et al. (2004) are also plotted (circle). The uncertainties in the observational luminosities and the effective 
temperatures are taken from Eggenberger et al. (2004). According to the NOS model, while $\alpha$ Cen A is very close to the end of 
its MS lifetime, $\alpha$ Cen B is at about half of its MS lifetime.  
The value of the mixing-length parameter found for $\alpha$ Cen A is greater than that for $\alpha$ Cen B. This result is in good agreement 
with the results given by Y{\i}ld{\i}z et al. (2006). 

The age of these models, 8.88 Gyr, also found by Miglio \& Montalban (2005), is almost the same as 
the age of the galactic thin disk (8.8 Gyr) 
obtained by del Peloso et al. (2006). Their age determination  is based on the abundance ratio of Th and Eu elements in 
26 stars including  $\alpha$ Cen.
	 
\subsection{Calibration of models with seismic constraints}
   \begin{figure}
\centerline{\psfig{figure=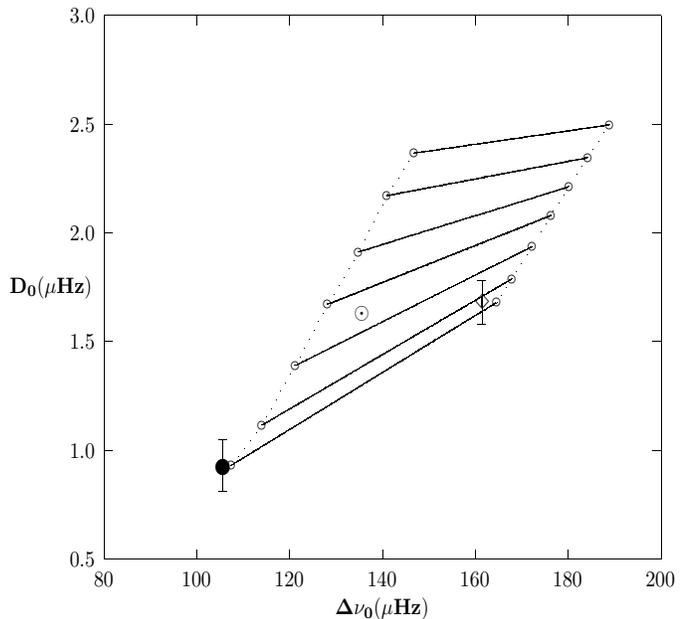,width=210bp,height=250bp}}
      \caption{Theoretical seismic HR diagram for $\alpha$ Cen A and B (SIS model). While the dotted lines
 show the evolutionary tracks of the stars, the solid lines show the isochrones. The solid line at the top 
corresponds nearly the ZAMS (0.1Gyr). The solid lines below it show 1, 2, 3, 4, 5 and 5.7 Gyr.
For comparison, the observed positions of the Sun($\odot$) and $\alpha$ Cen A (filled circle) and B (diamond) are also plotted. }

      {\label{f1.15}}
   \end{figure}

{
As discussed in the previous subsection, the small separations of NOS models of $\alpha$ Cen A and B are 
significantly smaller than the observed values.  
}
Therefore, models satisfying the seismic constraints 
must have a smaller age. However, models of $\alpha$ Cen A and B require a high value for age, such as 8.88 Gyr, in order to fit 
the observed difference between the luminosities of $\alpha$ Cen A and B. Because there is no simultaneous solution for 
$L_{\rm A}$, $L_{\rm B}$ and the seismic constraints, we therefore take into account the luminosity of one component only.
If we leave aside the luminosity of $\alpha$ Cen B and use all the remaining constraints including seismic measurements, 
we find a significantly smaller age for the system (SIS model) than predicted by NOS models. The SIS models 
of $\alpha$ Cen A and B are 
given in the third and fourth rows of Table 2, respectively. In contrast to the NOS models,
the value of the mixing-length parameter found for $\alpha$ Cen A is smaller than that of $\alpha$ Cen B.
However, the reason of this result could be that the luminosity of the SIS model for $\alpha$ Cen B is 
15 per cent greater than the observed luminosity.

{
If this excess is a result of observational uncertainty about the mass of  $\alpha$ Cen B,
then its required mass is 
}
0.914 $M_\odot$. The properties of the 
model with this mass
are given in the fifth row of Table 2. The mixing-length parameter of this model is less than 
the corresponding model of  $\alpha$ Cen A.

Because the abundance determination of stars is not an issue that has been pursued exhaustively, we also construct models for  $\alpha$ Cen A and B with 
two different $Z$. For $Z=0.03$ (sixth and seventh rows in Table 2) and $Z=0.023$ (eighth and ninth rows), 
putting aside the differences between the initial hydrogen abundances, we find very similar models. In order to see 
the effect of the heavy element mixture, we also obtain models with the recent solar heavy element mixture (AGS2005) in place of 
that of FG2001.



In Fig. 3, the time run of the models of $\alpha$ Cen A and B (SIS models) are plotted in a seismic HR diagram. The filled circle 
and diamond  
show $\alpha$ Cen A and B, respectively, and the Sun is represented by $\odot$. While the dotted lines mark the evolutionary tracks,
the solid lines show the isochrones. The isochrone line at the top is for 0.1 Gyr and the next line is for 1 Gyr. 
The increment between the other isochrone lines is 1 Gyr, except the last one. The isochrone line at the bottom is for 
5.7 Gyr. 

\subsection{On the alternative seismic HR diagrams}
{
Seismic and non-seismic constraints impose very different ages on the models of $\alpha$ Cen A and B.
For the small separation between the frequencies, however, 
oxburgh \& Vorontsov (2003) and Mazumdar (2005) propose 
alternative expressions for 
a better representation of physical conditions in central regions of the late-type stars.
Roxburgh \& Vorontsov (2003), for example, propose to use 
\begin{equation}
d_{01}(n)=\frac{1}{8}(\nu_{n-1,0}-4\nu_{n-1,1}+6\nu_{n,0}-4\nu_{n,1}+\nu_{n+1,0})
\end{equation}
in place of  $D_0$ (or $\delta \nu_{02}$). They  also confirm that 
the ratio of small to large separations ($r_{01}=d_{01}(n)i/\Delta \nu_{nl}$) for acoustic oscillations 
is more sensitive to physical
conditions in the stellar core. From the observed seismic data of $\alpha$ Cen A and B, we find that 
$r_{01}$ is 0.018 and 0.025, respectively. The uncertainty in $ d_{01}(n)$ is, in principle, the same as 
for the small separations. Then, the uncertainty in $ r_{01}$ is 0.007 for $\alpha$ Cen A and 0.004 for $\alpha$ Cen B. 
For the SIS model, we compute the ratio $r_{01}=d_{01}(n)i/\Delta \nu_{nl}$: $r_{01}=0.045$ for $\alpha$ Cen A and 
$r_{01}=0.012$ for $\alpha$ Cen B. Very similar results are also obtained for the other models listed in Table 2. 
While the value of $r_{01}$ for the model of $\alpha$ Cen A is significantly larger than the observed $r_{01}$ and does not show an
explicit time dependence, the observed $r_{01}$ of 
$\alpha$ Cen B is close to the zero-age main-sequence (ZAMS) value of the SIS model rather than the value at 5.7 Gyr.    
}
\subsection{Variability of the mixing-length parameter with time}
   \begin{figure}
\centerline{\psfig{figure=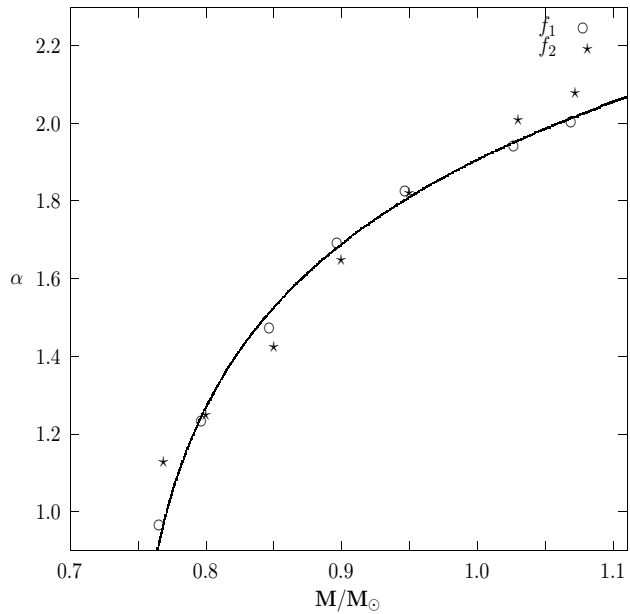,width=210bp,height=250bp}}
      \caption{The solid line represents the stellar mass dependence of the mixing-length parameter given by Y{\i}ld{\i}z et al. (2006) for the Hyades stars. 
               We try to derive more general expressions for $\alpha$ in terms of the structure of the convective zones rather than               the stellar mass. For two typical cases, we plot $f_1$ and $f_2$ as functions of some physical quantities of the convective zone (such as density and temperature at the base of convective zone and at the surface). See the text for the definitions of $f_1$ and $f_2$.}
              {\label{f1.12}}
   \end{figure}
   \begin{figure}
\centerline{\psfig{figure=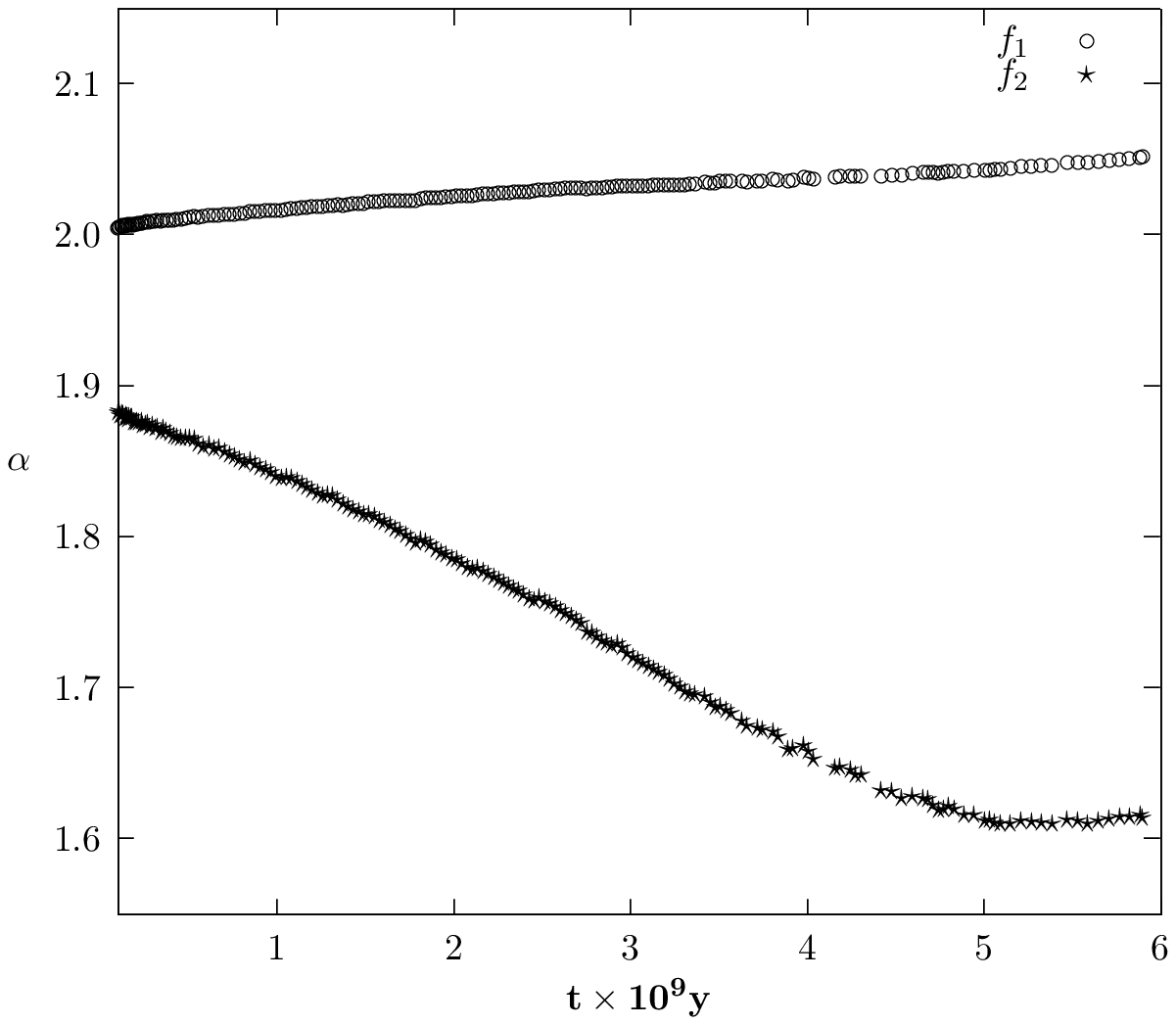,width=210bp,height=250bp}}
      \caption{The time dependence of the mixing-length parameter for a Cen A predicted by $f_1$ (circles) and $f_2$ (stars). }
              {\label{f1.13}}
   \end{figure}
   \begin{figure}
\centerline{\psfig{figure=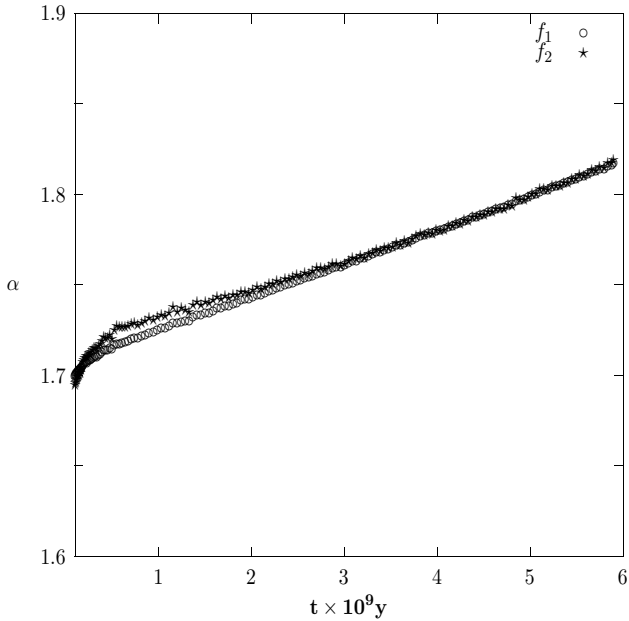,width=210bp,height=250bp}}
      \caption{The time dependence of the mixing-length parameter for a Cen B predicted by $f_1$ (circles)and $f_2$ (stars). }
              {\label{f1.14}}
   \end{figure}

As stated above, Y{\i}ld{\i}z et al. (2006) have confirmed a relationship between the mixing-length parameter for convection and 
the stellar mass from the late-type stars of the Hyades open cluster. According to this relationship, for $\alpha$ Cen A and B,
 $\alpha_{\rm A}=2.06 $ and $\alpha_{\rm B}=1.78$. A similar relationship is found for the models
of $\alpha$ Cen A and  B fitted into the non-seismic constraints of the system.  
The mixing-length parameters for these
models are $\alpha_{\rm A}=2.10 $ and $\alpha_{\rm B}=1.90$.
These parameters are qualitatively in very good agreement with the results of Y{\i}ld{\i}z et al. (2006) from the binaries of the Hyades open cluster:
the higher the mass, the greater the mixing-length parameter (case 1). 

However, this relationship may have different forms as the stars evolve away 
from the ZAMS (see below). For all models satisfying the seismic constraints, for example,
{ contrary}  to case 1, $\alpha_{\rm A}$ is less than $\alpha_{\rm B}$ (case 2). In case 2, at least for
 $\alpha$ Cen A, $\alpha$ may be a decreasing function of time. Neither case 1 nor case 2 gives the same values for 
$\alpha_{\rm A}$ and $\alpha_{\rm B}$. This means that $\alpha$ is not a simple function of stellar mass and
the mass dependence found by Y{\i}ld{\i}z et al. (2006) (see their figure 4) is a special case of a general situation about the ZAMS. 
So, we describe $\alpha$ as a function of some 
quantities pertaining to { the} convective zone of each model rather than stellar mass. In Fig. 4, 
we plot two such functions, $f_1$ (circles) and $f_2$ (stars), with respect to stellar mass:
\begin{equation}
f_1=2.5-{\rho _{\rm bcz}}{\left(\frac{2.7}{T_{6,\rm bcz}}\right)^4}-{\rho _{\rm ph}}{\left(\frac{1.4}{T_{5,\rm ph}}\right)^4},    
\end{equation}
\begin{equation}
f_2=\frac{3.25}{\rho _{\rm  bcz}^{0.8}}\left(\frac{M_{\rm bcz}}{r_{\rm bcz}^2}\right)^{0.5}{\left(\frac{T_{6,\rm bcz}}{3.8}\right)^4}   
\end{equation}
Here, the subscripts bcz and ph mean that the corresponding quantity pertain to the base of convective zone and
the photosphere, respectively. $r_{\rm bcz}$ is the radius of the base of the convective zone and $M_{\rm bcz}$ is the mass inside the sphere
with this radius. $r_{\rm bcz}$  and $M_{\rm bcz}$ are in solar units.
The solid line in Fig. 4 represents the fitting curve $\alpha (M)= 9.19 (M/M_\odot-0.74)^{0.053}-6.65$ derived by
Y{\i}ld{\i}z et al. (2006). The three functions predict very similar values of $\alpha$ for a given mass. However, 
there are very important differences between these three functions. While
$\alpha (M)$ is constant in time, $f_1$ and $f_2$ are implicitly functions of time because the structure of 
the convective zone changes in time.

In Fig. 5, $f_1$ (circles) and $f_2$ (stars) are plotted as a function of time for  $\alpha$ Cen A (SISZ23 model). While 
 $f_1$ is an increasing function of time,  $f_2$ is a decreasing function. We also see how $f_1$ and $f_2$ 
vary in time for  $\alpha$ Cen B. In Fig. 6, we see that both $f_1$ (circles) and $f_2$ (stars) are increasing functions of time
and predict almost the same value for  $\alpha$ Cen B for a given time. 

\section{Conclusions}


The masses and radii of the components of $\alpha$ Cen, which are essential quantities for stellar modelling, 
are accurately determined from the observations. Using the photometric measurements of Bessel (1990) and 
the tables for colours  and bolometric corrections of LCB1998, we find the absolute magnitudes of $\alpha$ Cen A and  B and 
then their luminosities and surface heavy element abundances: $L_{ A}=1.544 {\rm L_\odot}$, $L_{ B}=0.507{\rm L_\odot}$,
$0.074$ dex and $0.125$ dex for the metallicities of $\alpha$ Cen A and B, respectively ( $Z_{\rm A}=0.0237$ and $Z_{\rm B}=0.0267$, 
taking the metallicity of the tables of colours and bolometric correction as Z=0.02).
This difference between the metal abundances of $\alpha$ Cen A and  B can be 
interpreted as the result of the diffusion process (settling), which is faster at the bottom of convective 
zone of $\alpha$ Cen A than $\alpha$ Cen B. 

The calibration of models of $\alpha$ Cen A and  B into these 
non-seismic constraints yields the age of the system as 8.88 Gyr. The mixing-length parameters for these 
models are $\alpha_{\rm A}=2.10 $ and $\alpha_{\rm B}=1.90$.
These values of $\alpha$ are qualitatively in very good agreement with the results of Y{\i}ld{\i}z et al. (2006) from the binaries of the Hyades open cluster:
The higher the mass, the greater the mixing-length parameter. However, this relationship may have different forms as the stars evolve away 
from the ZAMS (see below). The seismic properties of these models are very different from the observations. 
By comparing small separations between the oscillations, we confirm that both models are much 
older than the observed stars. 
%

The reason for such a great age for the system is that the observed luminosity of 
$\alpha$ Cen A 
is much greater than that of $\alpha$ Cen B 
according to their masses.
Because $\alpha$ Cen A evolves faster than $\alpha$ Cen B, an old age is required for a simultaneous agreement
between the models and observations. However, we should also question the  accuracy of the observed values 
(e.g. the radii; see below).

Recently, del Peloso et al. (2006) determined the age of the galactic thin disc using Th/Eu cosmochronology from the 
analysis of selected stars, including the components of $\alpha$ Cen, as 8.8 Gyr. This age is very close to the age  
we found from the models of $\alpha$ Cen A and  B for non-seismic constraints.    

According to seismic properties of its components, the $\alpha$ Cen system has a much smaller age than 8.88 Gyr; in fact, it is about 5.6-5.9 Gyr.
With such an age, however, the luminosities of $\alpha$ Cen A and B can not simultaneously be fitted into
the observed luminosities. Therefore, there is no constraint on luminosity of $\alpha$ Cen B in 
the calibration of these models. Because of this strategy, the model luminosities of $\alpha$ Cen B 
based on the seismic properties are 15 percent larger than the observed luminosity.  
To remove this discrepancy, either the mass of $\alpha$ Cen B can be reduced to 0.914 $M_\odot$ or 
its radius can be 7 per cent higher than the observed radius in Kervella et al. (2003).
However, the discrepancy may be a result of some processes, such as rotational mixing, which are not involved in 
the model computations. 

For all the models that also satisfy the seismic properties, in contrast to the models for non-seismic 
constraints, $\alpha_{\rm B}$ is greater than $\alpha_{\rm A}$. If this is the case, then the mixing-length
parameter is not a simple function of stellar mass (Y{\i}ld{\i}z et al. 2006) at all evolutionary phases, 
but { is} also a function of time. Therefore, we derive two expressions ($f_1$ and $f_2$; see Section 3.4) based on the structure of 
the convective zone
of a given model and also compatible with the results in Y{\i}ld{\i}z et al. (2006) near the ZAMS. While one of the expressions
gives $\alpha_{\rm A}$ as an increasing function of time, the other gives it as a decreasing function. Meanwhile,
according to both expressions, $\alpha_{\rm B}$ increases. Such expressions may also 
explain why the mixing-length parameter varies for stars from phase to phase (see, for example, Ferraro et al. 2006)   

We also compute small separations between the oscillation frequencies from the alternative expressions given 
by Roxburgh \& Vorontsov (2003) for models and the observed seismic data. In this case, we obtain more 
complicated situation than in the classical seismic HR diagram. 
{
The small separation of models for $\alpha$ Cen A computed from 
the expression given by Roxburgh \& Vorontsov (2003) shows no explicit time dependence.
}
\section*{Acknowledgments}

J. Christensen-Dalsgaard is acknowledged for providing his adiabatic pulsation code. I also thank 
Ay\c{s}e Lahur K{\i}rtun\c{c} for { her}  suggestions which improved the language of the paper.
This work is supported by the Scientific and 
Technological Research Council of Turkey (T\"UB\.ITAK).

\def \apj#1#2{ApJ,~{#1}, #2}
\def \aj#1#2{AJ,~{#1}, #2}
\def \astroa#1{astro-ph/~{#1}}
\def \pr#1#2{Phys.~Rev.,~{#1}, #2}
\def \prt#1#2{Phys.~Rep.,~{#1}, #2}
\def \rmp#1#2{Rev. Mod. Phys.,~{#1}, #2}
\def \pt#1#2{Phys.~Today.,~{#1}, #2}
\def \pra#1#2{Phys.~Rev.,~{ A}~{#1}, #2}
\def \asap#1#2{A\&A,~{#1}, #2}
\def \aandar#1#2{A\&AR,~{#1}, #2}
\def \apss#1#2{~Ap\&SS,~{{#1}}, #2}
\def \asaps#1#2{A\&AS,~{#1}, #2}
\def \arasap#1#2{ARA\&A,~{#1}, #2}
\def \pf#1#2{Phys.~~Fluids,~{#1}, #2}
\def \apjs#1#2{ApJS,~{#1}, #2}
\def \pasj#1#2{PASJ,~{#1}, #2}
\def \mnras#1#2{MNRAS,~{#1}, #2}
\def \ibvs#1{IBVS,~No.~{#1}}

\label{lastpage}

\end{document}